\documentclass[journal]{IEEEtran}

\ifCLASSINFOpdf
\else
\fi

\usepackage{amsmath}
\usepackage[T1]{fontenc}
\usepackage[utf8]{inputenc}
\usepackage{graphicx}
\usepackage{xcolor}
\usepackage{tgtermes}
\usepackage{amsmath,amssymb,amsthm,textcomp}
\usepackage{enumerate}
\usepackage{multicol}
\usepackage{tikz}
\usepackage{pdfpages}
\graphicspath{{./figures/}}

\begin{document}

\title{Service Allocation in a Mobile Fog Infrastructure under Availability and QoS Constraints}
\author{Nader~Daneshfar, Nikolaos~Pappas, Valentin~Polishchuk and~Vangelis~Angelakis
\thanks{N. Daneshfar, N. Pappas, Valentin Polishchuk and V. Angelakis are with Department of Science and Technology, Link\"{o}ping University, Norrk\"{o}ping, Sweden. Email: \{nader.daneshfar, nikolaos.pappas, valentin.polishchuk, vangelis.angelakis\}@liu.se.}
}

\maketitle

\begin{abstract}
	The next generation of mobile networks, namely 5G, and the Internet of Things (IoT) have brought a large number of delay sensitive services. In this context Cloud services are migrating to the edge of the networks to reduce latency. The notion of Fog computing, where the edge plays an active role in the execution of services, comes to meet the need for the stringent requirements. Thus, it becomes of a high importance to elegantly formulate and optimize this problem of mapping demand to supply. This work does exactly that, taking into account two key aspects of a service allocation problem in the Fog, namely modeling cost of executing a given set of services, and the randomness of resources availability, which may come from pre-existing load or server mobility. We introduce an integer optimization formulation to minimize the total cost under a guarantee of service execution despite the uncertainty of resources availability.
	
\end{abstract}

%\let\thefootnote\relax\footnote{This project has received funding from the European Union’s Horizon 2020 research and innovation programme under the Marie Sklodowska-Curie grant agreement No "642743" (WiVi-2020).}


\section{Introduction}

Driven primarily by requirements of the Internet of Things (IoT), Fog networking has been introduced as an architecture running services at the edge of the network \cite{Bonomi:2012:FCR:2342509.2342513}. Fog can enhance distributed computing, management, control, storage and networking by providing such services at the edge of the network \cite{ServiceHandlingCombinedFogCloud}. Comparing Fog against the Cloud we find different key features that can be categorized into three main parts as \textit{Storage}, \textit{Computation} and \textit{Network Communication and Management}.
Any operating system is highly dependent on data handling and processing. In this respect, applications either have their own capability for storing data or utilize a remote resource upon request. Fog can introduce temporary storage at the edge of network to localize the file storage management.
Fog Radio Access Network (F-RAN) architecture has been introduced to alleviate merging existing technologies and combining the benefits of both edge and Cloud processing \cite{FogComputingRANChallenges}. Finally, considering the size of vehicular fleets in urban and rural areas, introduces an
underutilized set of resources that offers considerable opportunities for networking innovation \cite{tvt}.

Fog is still a relatively young term, with different definitions, architectures and scopes. Even the term Fog networking is often interchangeable with Fog computing. Fog computing is defined as a very large number of interconnected heterogeneous and decentralized devices that have the ability to communicate and cooperate with each other and the existing network to facilitate performing tasks without the intervention of third parties \cite{FogIotOverview}.

Similar definitions to Fog (especially in the computing scheme) exist in the current literature such as Mobile Edge Computing (MEC) and Mobile Cloud Computing (MCC). The former is focused on bringing more computational power to the edge of the network where users demand higher level of computational power \cite{IoTMECArchitecture,IoTMECCloudComputing}. The latter describes a method where both computational task and storage occur at a remote place with respect to the mobile node \cite{IoTMCC}. Additionally, authors in \cite{HeterogeneousResourcesIoannis} propose a solution to address the allocation of heterogeneous demands into available resources.

\subsection{Motivation}
The potential mobility of the Fog networking elements is highly relevant, since user mobile devices have considerable computing and storing capabilities. Considering mobility to network elements can be considered to increase the uncertainty or reduce reliability in their role as service providers, increasing the importance of investigating the effect of this phenomenon on the users' QoS. At the same time, the heterogeneity of the servers in a Fog infrastructure gives rise to a host of issues that are still not well investigated.

\subsection{Contribution}
We introduce a Integer Programming formulation that minimizes the total cost of providing services while allocating demands to available resources. In the formulation we associate each servers with a ``probability of availability'' that captures potentially mobility and formulate its effect on the QoS of each user's demand. On this note, the QoS per each user's demand is then assured by allowing duplication. We show that the problem we formulate is computationally hard and we give numerical results that provide fundamental understanding of the effect of service and servers set compositions, and of the availability of the latter to the solution cost.

\section{Problem Definition and Notation}
The problem we consider amounts to deciding for a set of users to which set of Fog servers each user should multicast their demand to meet a predefined threshold for service reliability. That is, we consider the problem from the perspective of a centralized controller (middleware) between the users and the Fog, having \textit{complete and correct} information about all the elements of the Fog infrastructure and users, but has a well known non-deterministic component of availability for each of the servers.

\begin{figure}[t]
\centering
\includegraphics[scale=0.25]{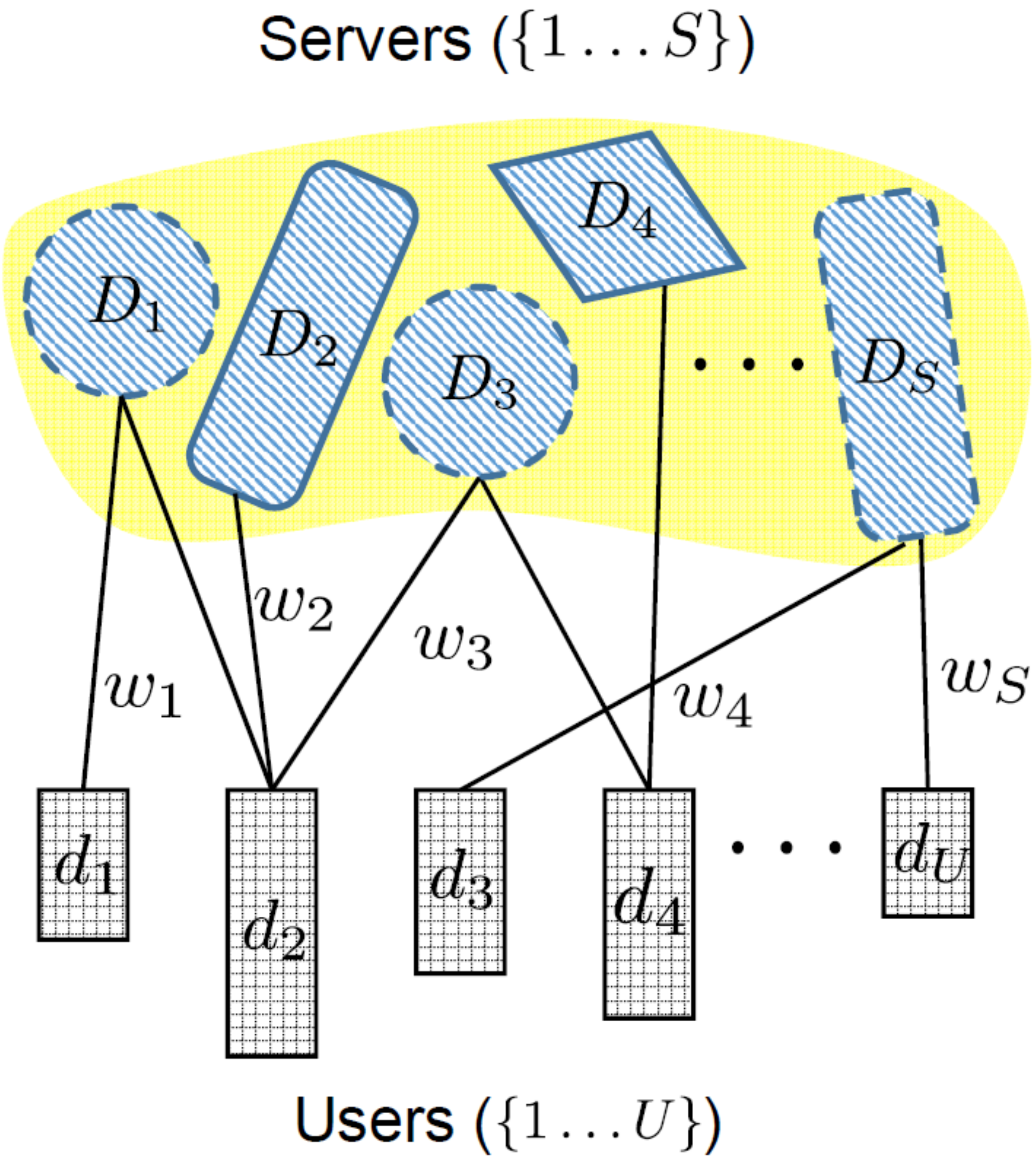}
\caption{A set $\mathcal{U}$ of users demand some service from a set $\mathcal{S}$ of Fog servers. The demand from a user to a server $s$ has a cost $w_s$.} 
\label{fig:model}
\end{figure}

Specifically, we consider that we are given a set $\mathcal{U}$ of users that each demands some service from a set $\mathcal{S}$ of Fog servers as depicted in Fig. \ref{fig:model}. The amount of demand of user $u\in \mathcal{U}$ is denoted $d_u$. Each user service demand can not be split for service by more than one servers. To combat the unknown server availability, we allow multicasting of the service demand to multiple servers. Flooding a user's demand (i.e., utilizing way too many servers) is avoided by introducing a bound $M\leq\left\vert\mathcal{S}\right\vert$, on the number of servers a single user can post the service to. Sending a unit of the demand to server $s\in \mathcal{S}$ has a \textit{cost} $w_s > 0$ and each user $u\in \mathcal{U}$ has a limited \textit{budget} $B_u > 0$ within which the total cost incurred by $u$'s service assignments must remain. A server's resources may be used  by multiple users as long as the total demand served by server $s$ should not exceed its capacity $D_s$.  A server $s\in \mathcal{S}$ is available with probability $p_s\in(0,1)$ and we assume that the availabilities of different servers are independent. Thus, successful completion of services cannot be guaranteed to the users. Each user $u\in \mathcal{U}$ however has an expressed  \textit{minimum service level} requirement $l_u\in(0,1)$ meaning the user wants to have its demand served with probability no lower than $l_u$. A summary of the notation used can be found in \tablename{ \ref{table:notations}}. We dub our problem as \textit{M-Fog Allocation} (for mobile Fog service allocation), or \textit{MFA} for short.

\begin{table}[!t]
\caption{Notations}
\label{table:notations}
\centering
\begin{tabular}{c|p{7cm}}
\hline
\bfseries Symbol & \bfseries Definition\\
\hline\hline
$\mathcal{U}$ & Set of users $\mathcal{U}=\{1,2,\dots,U\}$\\
$\mathcal{S}$ & Set of servers $\mathcal{S}=\{1,2,\dots,S\}$\\
$\left\vert\mathcal{S}\right\vert$ & Cardinality of $\mathcal{S}$\\  
$d_u$ & Demand value of user $u$\\
$l_u$ & Minimum service level requirement of user $u$\\
$B_u$ & Budget of user $u$\\
$M$ & Maximum number of servers a user is allowed to multicast a service to\\
$p_s$ & Probability of availability of server $s$\\
$w_s$ & Cost of sending a unit of demand to server $s$\\
$D_s$ & Maximum amount of total demand that can be assigned to server $s$\\
\hline
\end{tabular}
\end{table}

A natural objective function is the total cost incurred by the users when multicasting their demand. For simplicity we assume each service has a cost of $w_s$. We emphasize that $w_s$ is the cost of \textit{sending} one unit of the demand to server $s$, which a user must pay \textit{irrespectively} of whether the server is there (and thus serves the user) or not.

This completes the description of the basic version of MFA. In further sections we consider extensions of the model -- a) a constraint that restricts overloading of the servers with excessive demands and b) an objective for maximizing the minimum achieved probability of service.

\section{Modeling MFA as an Integer Program}
We formulate MFA as an Integer Program (IP) whose decision variables $x_{us},u\in \mathcal{U},s\in \mathcal{S}$ indicate whether user $u$ includes server $s$ into the set of servers to which it multicasts its demand:
\begin{equation*} x_{us}=\left\{
\begin{array}{cl}
1&\textrm{if $u$ sends to $s$}\\
0&\textrm{otherwise}\\
\end{array}
\right.
\end{equation*}

The objective is to minimize the total cost:
\begin{equation}\label{eq:objFunction}
\sum_{u \in \mathcal{U}}\sum_{s \in \mathcal{S}}d_uw_sx_{us},
\end{equation}

Such that the following constraints are met:

\begin{equation}\label{eq:userConnection}
\sum_{s \in \mathcal{S}} {x_{us}} \leq M, \quad  \forall u\in \mathcal{U},
\end{equation}
%\vspace{-15pt}

\begin{equation}\label{eq:userBudget}
\sum_{s \in \mathcal{S}} {d_uw_s}{x_{us}} \leq B_{u}, \quad  \forall u\in \mathcal{U},
\end{equation}
%\vspace{-15pt}

\begin{equation}\label{eq:serverCapacity}
\sum_{u \in \mathcal{U}} d_{u}{x_{us}} \leq D_{s}, \quad \forall s\in \mathcal{S},
\end{equation}
%\vspace{-15pt}

\begin{equation}\label{eq:qos}
\sum_{s \in \mathcal{S}}x_{us}\ln(1-p_s)\le\ln(1-l_{u}), \quad \forall u\in \mathcal{U}.
\end{equation}

With the constraint in (\ref{eq:userConnection}) we limit the number of servers each user can use. Since each user is coupled with a budget (e.g. battery, quota, etc.) the constraint in (\ref{eq:userBudget}) ensures no user exceeds this. On the other hand, the model takes into account the capacity of the servers that the Fog consists of. The constraint imposed by (\ref{eq:serverCapacity}) limits the total amount of service requests to be sent from various users to server set in Fog. The set of constraints in (\ref{eq:qos}) offer users their requested QoS Level. Specifically, we assume that in our model QoS is captured as a maximum probability of service failure, i.e. the probability that the user's request will not be served by any of the assigned servers. Observe then that we can rewrite the left hand side as $\sum_{s:x_{us}=1}\ln(1-p_s)$ and then taking an exponent on both sides we get $\Pi_{s:x_{us}=1}(1-p_s)\le1-l_{u}$, where on the left side is now the desired probability.

A key property of our model is its flexibility to capture different network tiers (e.g. Cloud, Fog, Mobile Edge, etc). While well-known Cloud based networks include a big server and use 1-to-many connections \cite{UserBasedCloud}, edge networks benefit from several edge elements (e.g. base stations, home routers, even portable devices, etc). Our model is designed in such a way that easily fits a wide range of network types adjusting the cost and availability parameters. For example, a Cloud server could be considered to have a high cost for services to be assigned to, but high availability probability, while a user device could be modeled with low cost but also low probability of availability.

\section{Problem Complexity}
We will show that the problem is NP-Complete by reduction from a well known problem. The 3-Partition problem amounts to deciding whether a given set of integers can be partitioned into triplets, such that the sum of the integers in each triple does not exceed a given bound. The MFA can be reduced from the 3-Partition as follows: The number of users $|\mathcal{U}|$ is equal to the number of integers in the 3-Partition instance, and the integers become the demands $d_u,u\in \mathcal{U}$ of the users; the capacity of each server is equal to the bound on the sum of integers in the triples, and the probability $p_s=1$ for each server $s$. Set $M=1$, the weights $w_s=0$ and set the budgets $b_u$ arbitrarily (since the costs are 0s, the budget is not an issue). Set $l_u=1$, so every user will have to send to a server. Since $M=1$, we have to decide, for each user, to which server to send. The 3-Partition instance is feasible if it is possible to send all users demands to the servers without violating the servers capacities. Note that although (\ref{eq:qos}) does not allow $p_s=1$, the proof holds because this is only for linearization purposes and does not affect the structure of the problem itself.

The above argument shows that even deciding the \textit{feasibility} of an instance of MFA is hard. It follows that no approximation to the objective function is possible to find in polynomial time (unless P=NP).

\section{Numerical Results}
In this section we present the results for sets of simulations that has been conducted using Matlab software integrated with Gurobi Optimizer solver.
Here we aim to asses the optimal total cost of serving all service requests of the system. This is conducted with different configurations of cost per unit of service request as well as the probability of availability for servers and the minimum QoS requirement by users.
The former test condition is aimed to study the effect of different distribution in the cost variable while the latter is investigating the trade-off between QoS request and resource availability. It is important to note that each user's available budget is ensured to be sufficiently adjusted, ensuring that they can send as many number of demand duplicates as required to achieve their desired quality of service. Additionally, all scenarios are performed with fixed number of servers having constant accumulated available resources in total.

%\vspace{-10pt}

\begin{figure}[htbp]
	\centering
	\includegraphics[width=.5\textwidth]{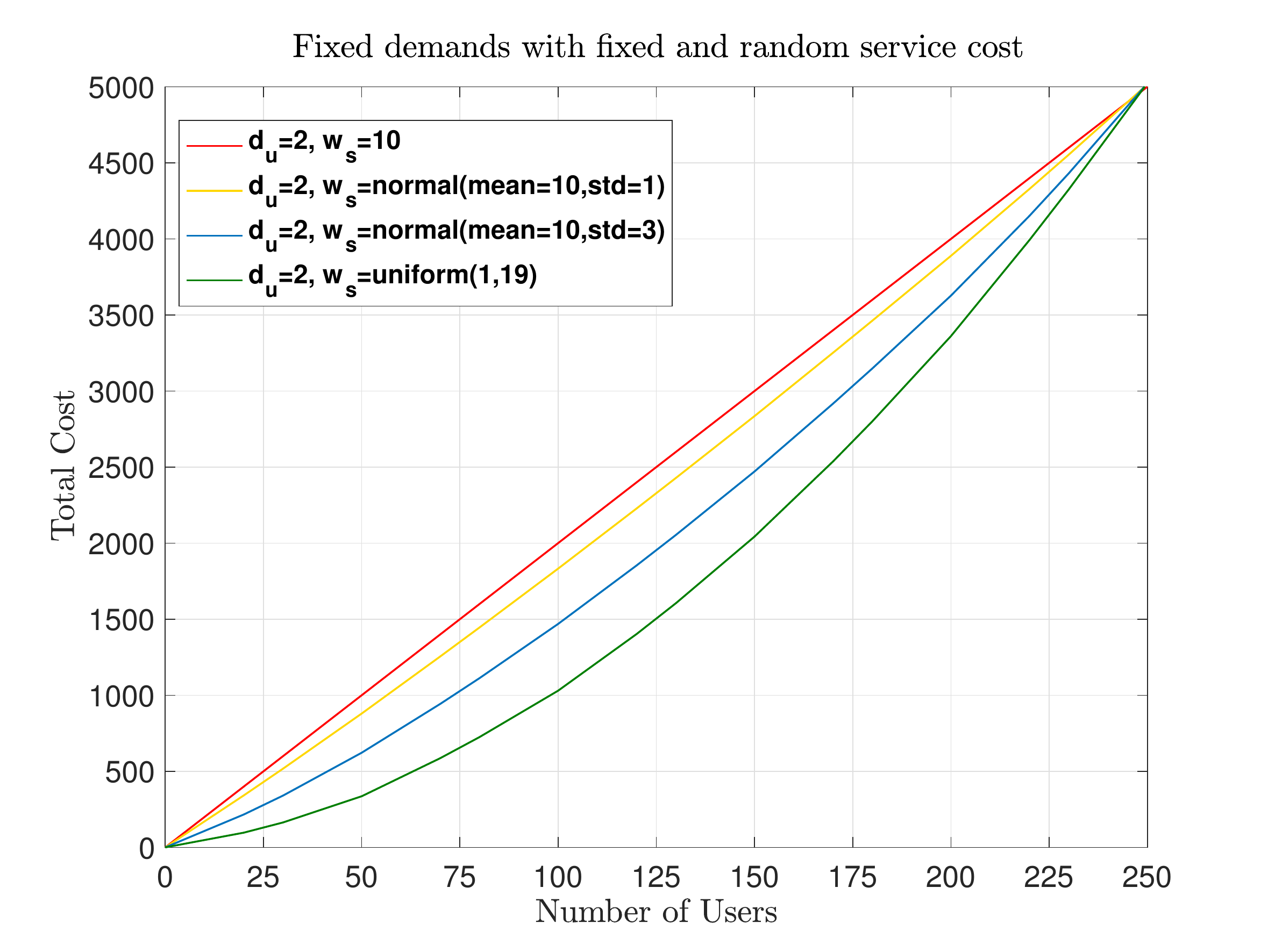}
	\caption{An instance of network deployment with 10 servers each having 50 units of services. Fixed $d_u=2$  and fixed $w_s=10$ units. Random  $w_s$ scenarios are set of uniformly generated random integers in the interval of $[1,19]$ and normally generated random values with $\mu=10$ and $\sigma=\{1,3\}$.}
	\label{fig:userVScost}
\end{figure}

The total cost is at its highest optimal value when the cost per unit of service ($w_s$) is fixed at a constant value and it increases linearly. It is caused by the fact that all servers provide services with identical cost thus, eliminating users to choose which server to send their service request. The optimal cost for a number of users decreases as the servers get more diverse with their available cost per unit of request. This phenomenon is clearly visible in Fig. \ref{fig:userVScost} especially when the system is not saturated. This is because as more diversity is implemented into server set, the users can send their request to the server with the most inexpensive resources.

%\vspace{-15pt}

\begin{figure}[htbp]
	\centering
	\includegraphics[width=.5\textwidth]{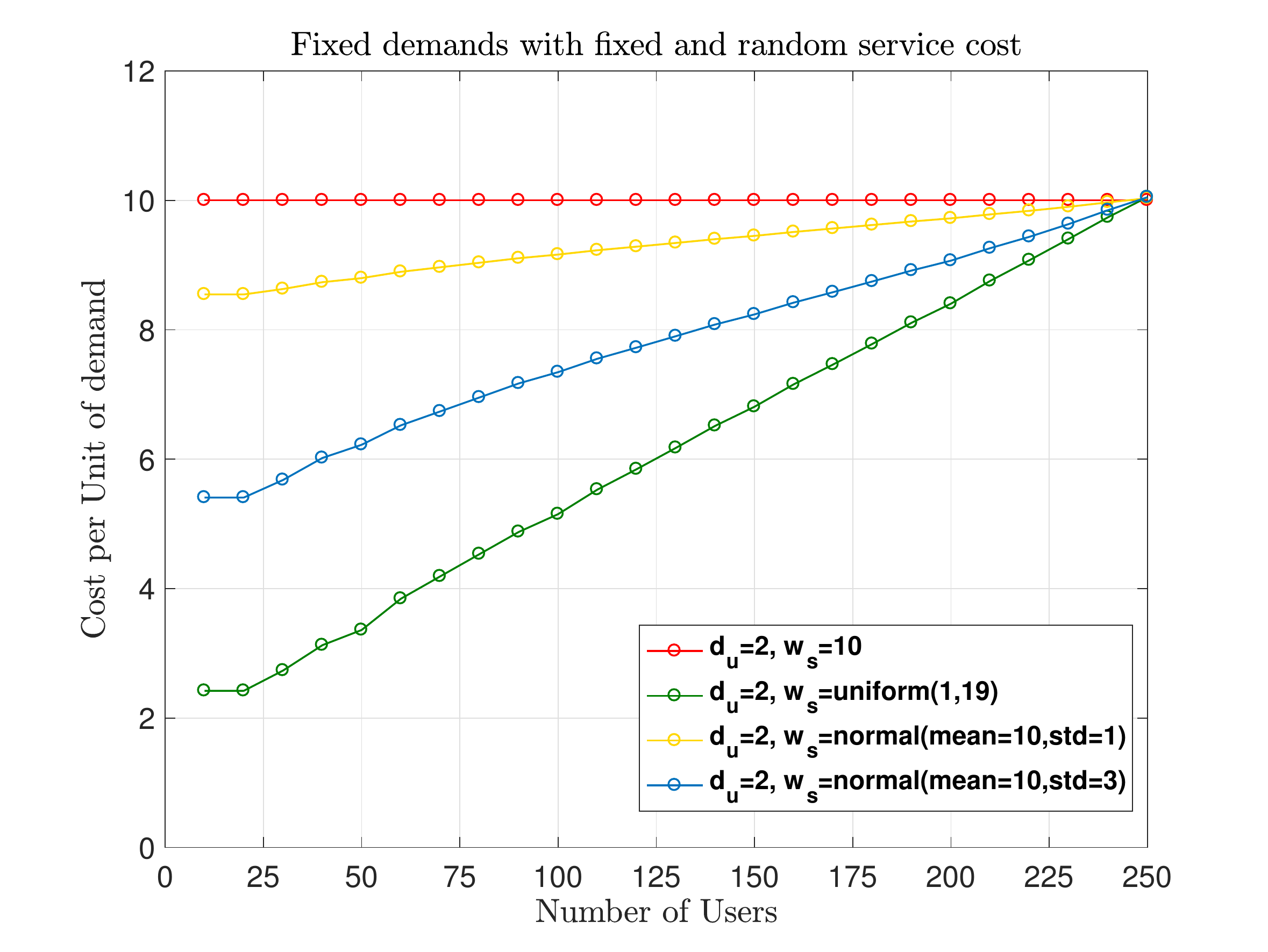}
	\caption{Total cost vs the total number of users with fixed demands of 2 units. Cost per unit of service is being distributed \textit{uniformly} and \textit{normally} with means equal to 10. Different slopes illustrated in the figure is in direct relation with the diversity of servers with various service cost.}
	\label{fig:CostPerUnit}
\end{figure}

The average cost per requesting a unit of demand from the server set ultimately converges to the scenario where the cost for requesting a unit of demand is fixed as the number of users increase to the point of saturating the network. The results depicted by Fig. \ref{fig:CostPerUnit} indicate that in all scenarios the initial average cost per unit of demand, despite having various values, is not changing for a short increment in the number of users. This is due to the fact that initially users' requests are directed to the available services on servers with the minimum cost per demand. Considering the total cost per unit of demand is divided among users, the average cost per unit of service then starts to incline as the servers begin to get occupied by the least costly to mostly costly fashion.

%\vspace{-10pt}

\begin{figure}[htbp]
	\centering
	\includegraphics[width=.5\textwidth]{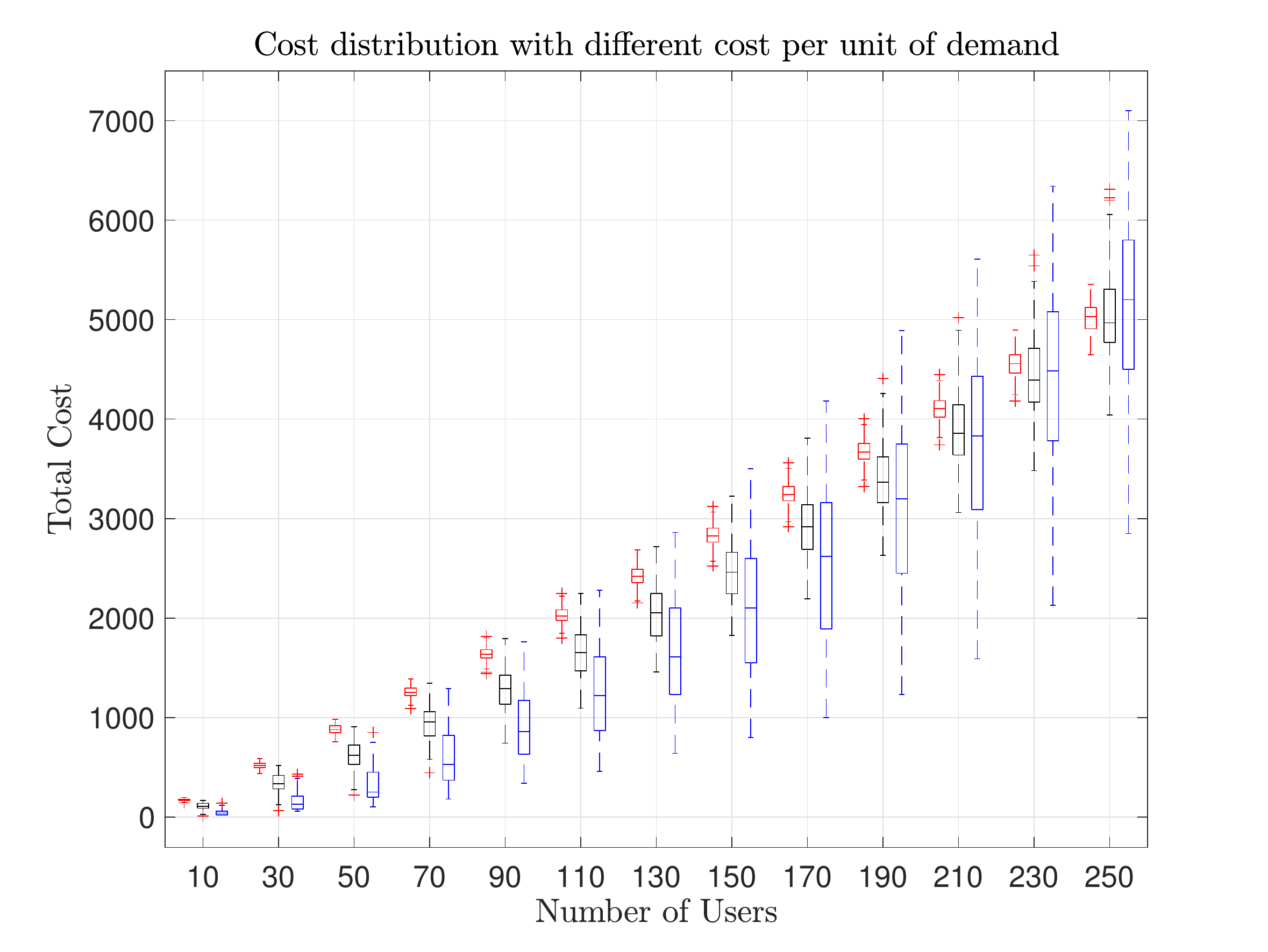}
	\caption{Total number of users and the respective total cost of providing services to them is shown here. The figure depicts the variation of total cost when the system consists of servers with uniform (blue) and normal (black and red) distribution of service cost. All distributions has a mean equal to 10. Deviations for normal distributions are 3 for former and 1 for the latter.}
	\label{fig:CostDistribution}
\end{figure}

Fluctuations in the total cost of covering all the service request of all the users is highly impacted by the distribution of servers in a scenario. This phenomena can be characterized as diversity of servers. As illustrated in Fig. \ref{fig:CostDistribution} there exists an explicit correlation when the number of servers, having various capacities, tend to have more diverse cost per unit of demand compared to each other to the case where the capacity is chosen of values with less difference around a mean value. The figure also provides an illustrative information for best and worst case conditions where the majority of servers with available resources have low and high cost per unit of the services they provide.

%\vspace{-10pt}

\begin{figure}[htbp]
	\centering
	\includegraphics[width=.5\textwidth]{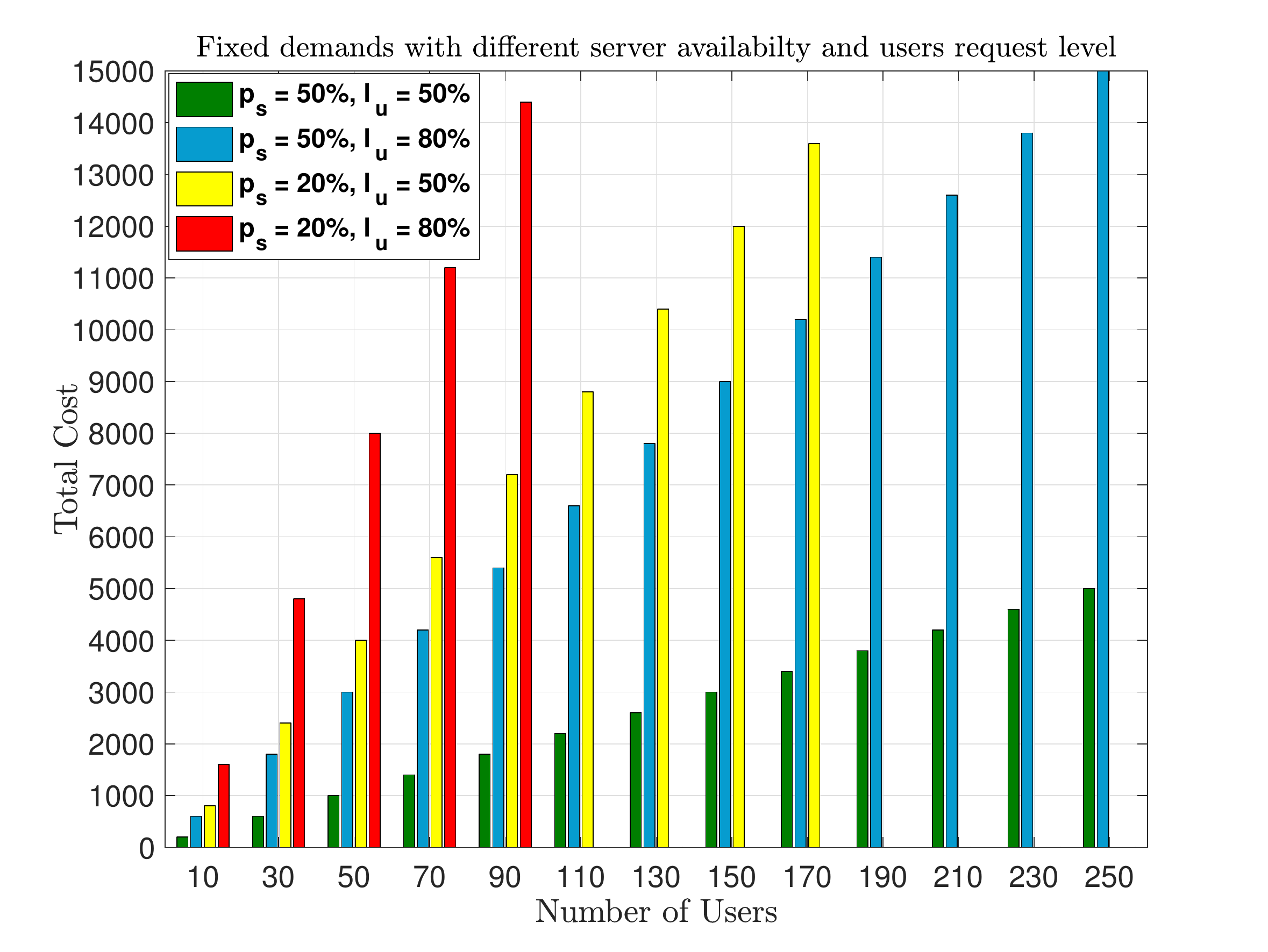}
	\caption{An example of implementation with fixed number of servers set to $S=30$ and 50 unit capacity. Different probability of availability for servers and minimum service requirements of each user is investigated, showing the overall cost of allocating services to users.}
	\label{fig:diffAvailability}
\end{figure}

In addition, as shown in Fig. \ref{fig:diffAvailability}, when $w_s$ is fixed, the total cost of serving all users' demands is in linear relation with the number of replications each user must have. This is to ensure that all users' required QoS are met. The bars reflect the fact that for higher QoS request, with a fixed server availability, more replications of service request are needed. It is remarkable that the multiplication factor for request replications is not constant with QoS increase in different server availabilities. This is due to the logarithmic behavior of the model. Moreover, optimum values for some scenarios (namely red and yellow) are discontinued because of the available resources and the fact that service replications extend beyond server set capacity to provided services.

\section{Conclusion}
This research aims to address mobility as one of the important characteristics of Fog networks. In this article, a model is introduced that takes the availability of services into account while optimizing the total service cost. The objective function of minimizing cost takes into account the amount of demand that is being requested by each user and also the cost per unit of service that is being provided by each server.

Multiple scenarios have been introduced and tested to evaluate the behavior of our model. Initially cost and demand variables are fixed to observe the system performance under extreme limits which the results act as a guideline for the other set of results. Three various randomization of per unit cost for services has been conducted. First a test for servers with uniformly distributed costs is done followed by two set of normally distributed cost per unit with different deviation to mimic the natural characteristics of real world scenarios. Furthermore, we have conducted another scenario where we target to investigate how our model acts in respect of providing the requested QoS from user.

The numerical results driven from scenarios outline the ability of our model to minimize total cost. They also successfully indicate the effect of various system variables such as the probability of availability of each server in respect with the minimum service required by each user and diversity in the server set regarding their service cost. Overall, users can achieve their requested level of QoS by making duplicates of their requests to multiple servers but it comes with a price and also saturates the whole network. On the other hand, our model indicates the direct relation of diversity in cost of services provided by servers and their quantity. As a Fog network tends to have more diverse servers, the further a system can reach its lowest optimized value in relation to its total cost of providing services.

\section*{Acknowledgment}
The research leading to these results has been funded from the European Union’s Horizon 2020 research and innovation programme under the Marie Sklodowska-Curie grant agreement No ``642743'' (WiVi-2020).

\bibliographystyle{IEEEtran}
\bibliography{IEEEabrv,mybibfile}

\begin{thebibliography}{10}
\providecommand{\url}[1]{#1}
\csname url@samestyle\endcsname
\providecommand{\newblock}{\relax}
\providecommand{\bibinfo}[2]{#2}
\providecommand{\BIBentrySTDinterwordspacing}{\spaceskip=0pt\relax}
\providecommand{\BIBentryALTinterwordstretchfactor}{4}
\providecommand{\BIBentryALTinterwordspacing}{\spaceskip=\fontdimen2\font plus
\BIBentryALTinterwordstretchfactor\fontdimen3\font minus
  \fontdimen4\font\relax}
\providecommand{\BIBforeignlanguage}[2]{{%
\expandafter\ifx\csname l@#1\endcsname\relax
\typeout{** WARNING: IEEEtran.bst: No hyphenation pattern has been}%
\typeout{** loaded for the language `#1'. Using the pattern for}%
\typeout{** the default language instead.}%
\else
\language=\csname l@#1\endcsname
\fi
#2}}
\providecommand{\BIBdecl}{\relax}
\BIBdecl

\bibitem{Bonomi:2012:FCR:2342509.2342513}
F.~Bonomi, R.~Milito, J.~Zhu, and S.~Addepalli, ``Fog computing and its role in
  the internet of things,'' in \emph{Proceedings of the First Edition of the
  MCC Workshop on Mobile Cloud Computing}.\hskip 1em plus 0.5em minus
  0.4em\relax New York, NY, USA: ACM, 2012, pp. 13--16.

\bibitem{ServiceHandlingCombinedFogCloud}
V.~B.~C. Souza, W.~Ramírez, X.~Masip-Bruin, E.~Marín-Tordera, G.~Ren, and
  G.~Tashakor, ``Handling service allocation in combined fog-cloud scenarios,''
  in \emph{IEEE International Conference on Communications (ICC)}, May 2016,
  pp. 1--5.

\bibitem{FogComputingRANChallenges}
M.~Peng, S.~Yan, K.~Zhang, and C.~Wang, ``Fog-computing-based radio access
  networks: issues and challenges,'' \emph{IEEE Network}, vol.~30, no.~4, pp.
  46--53, July 2016.

\bibitem{tvt}
X.~Hou, Y.~Li, M.~Chen, D.~Wu, D.~Jin, and S.~Chen, ``Vehicular fog computing:
  A viewpoint of vehicles as the infrastructures,'' \emph{IEEE Transactions on
  Vehicular Technology}, vol.~65, no.~6, pp. 3860--3873, June 2016.

\bibitem{FogIotOverview}
M.~Chiang and T.~Zhang, ``Fog and {IoT}: An {Overview} of {Research}
  {Opportunities},'' \emph{IEEE Internet of Things Journal}, vol.~3, no.~6, pp.
  854--864, Dec 2016.

\bibitem{IoTMECArchitecture}
D.~Sabella, A.~Vaillant, P.~Kuure, U.~Rauschenbach, and F.~Giust, ``Mobile-edge
  computing architecture: The role of mec in the internet of things,''
  \emph{IEEE Consumer Electronics Magazine}, vol.~5, no.~4, pp. 84--91, Oct
  2016.

\bibitem{IoTMECCloudComputing}
P.~Corcoran and S.~K. Datta, ``Mobile-edge computing and the internet of things
  for consumers: Extending cloud computing and services to the edge of the
  network,'' \emph{IEEE Consumer Electronics Magazine}, vol.~5, no.~4, pp.
  73--74, Oct 2016.

\bibitem{IoTMCC}
Y.~Liu, M.~J. Lee, and Y.~Zheng, ``Adaptive multi-resource allocation for
  cloudlet-based mobile cloud computing system,'' \emph{IEEE Transactions on
  Mobile Computing}, vol.~15, no.~10, pp. 2398--2410, Oct 2016.

\bibitem{HeterogeneousResourcesIoannis}
V.~Angelakis, I.~Avgouleas, N.~Pappas, E.~Fitzgerald, and D.~Yuan, ``Allocation
  of heterogeneous resources of an iot device to flexible services,''
  \emph{IEEE Internet of Things Journal}, vol.~3, no.~5, pp. 691--700, Oct
  2016.

\bibitem{UserBasedCloud}
X.~Yingying, A.~Naixiang, H.~Dan, Z.~Yongxiang, and C.~Changjia, ``Advanced
  user-based interaction model in cloud,'' \emph{China Communications},
  vol.~10, no.~4, pp. 126--134, April 2013.

\end{thebibliography}

\end{document}